\documentclass[twocolumn,superscriptaddress,floatfix,preprintnumbers, nofootinbib,hyperref]{revtex4-2} 
\pdfoutput=1
\usepackage[colorlinks=true,breaklinks=true]{hyperref}
\usepackage[normalem]{ulem}
\usepackage[utf8]{inputenc}
\hypersetup{allcolors=[rgb]{0.0 0.0 0.6},linkcolor=[rgb]{0.75 0.05 0.05}}
\usepackage{amsmath,amssymb}
\usepackage{epsfig}  
\usepackage{graphicx}   
\usepackage{slashed}       
\usepackage{tikz}
\usepackage{tikz-feynman}
\usepackage{url}
\usepackage{color}
\usepackage{multirow}
\usepackage{comment}
\usepackage{amssymb}
\usepackage{xcolor}

\allowdisplaybreaks

\bibpunct{[}{]}{,}{n}{}{,}

\pgfarrowsdeclare{X}{X}
{
  \pgfutil@tempdima=0.3pt%
  \advance\pgfutil@tempdima by.25\pgflinewidth%
  \pgfutil@tempdimb=5.5\pgfutil@tempdima\advance\pgfutil@tempdimb by.5\pgflinewidth%
  \pgfarrowsleftextend{+-\pgfutil@tempdimb}
  \pgfarrowsrightextend{0pt}
}
{
  \pgfutil@tempdima=0.3pt%
  \advance\pgfutil@tempdima by.25\pgflinewidth%
  \pgfsetdash{}{+0pt}
  \pgfsetroundcap
  \pgfsetmiterjoin
  \pgfpathmoveto{\pgfqpoint{-5.5\pgfutil@tempdima}{-6\pgfutil@tempdima}}
  \pgfpathlineto{\pgfqpoint{5.5\pgfutil@tempdima}{6\pgfutil@tempdima}}
  \pgfpathmoveto{\pgfqpoint{-5.5\pgfutil@tempdima}{6\pgfutil@tempdima}}
  \pgfpathlineto{\pgfqpoint{5.5\pgfutil@tempdima}{-6\pgfutil@tempdima}}
  \pgfusepathqstroke 
}
\makeatother

\newcommand{\Tr}{{\rm{Tr}}}
\newcommand{\Ree}{{\rm{Re}}}

\usepackage{listings}

\definecolor{lightgray}{gray}{0.95}
\definecolor{codegreen}{rgb}{0,0.6,0}
\definecolor{codegray}{rgb}{0.5,0.5,0.5}
\definecolor{codepurple}{rgb}{0.58,0,0.82}
\definecolor{backcolour}{rgb}{0.95,0.95,0.92}

\lstdefinestyle{mystyle}{
    frame=single,
    backgroundcolor=\color{lightgray},   
    keywordstyle=\color{magenta},
    numberstyle=\tiny\color{codegray},
    stringstyle=\color{blue},
    basicstyle=\ttfamily\footnotesize,
    breakatwhitespace=false,         
    breaklines=true,                 
    captionpos=b,                    
    keepspaces=true,                 
    numbers=left,                    
    numbersep=5pt,                  
    showspaces=false,                
    showstringspaces=false,
    showtabs=false,                  
    tabsize=2
}

\begin{document}
\title{Systematic derivation of perturbative unitarity bounds for any models}

\author{Nico Benincasa}
\email{nico.benincasa@kbfi.ee}
\affiliation{School of Physics, University of Electronic Science and Technology of China, 611731 Chengdu, China}

\smallskip

\begin{abstract}
In this letter, we provide a simple algorithm, \texttt{anyPUB}, to systematically derive the $2\rightarrow 2$ scattering matrix in the high-energy limit for any kind of models, irrespective of their gauge group or their field representation. After computing the eigenvalues analytically and/or numerically from this matrix, we impose perturbative unitarity bounds on them. We tested our method on various models and validated the results against the literature. Finally, as a concrete application of our approach, we discuss the case of the minimal left-right symmetric model and derive, for the first time, the perturbative unitarity constraints in the Pati-Salam model. 
\end{abstract}

\maketitle

\section{Introduction}
\label{sec:intro}


Although the discovery of the Higgs boson at the LHC~\cite{ATLAS:2012yve, CMS:2012qbp}, completed the Standard Model (SM), the latter cannot account for several key phenomena, such as dark matter, neutrino masses, or the baryon asymmetry of the Universe. These shortcomings motivate the study of physics beyond the Standard Model (BSM). Many of these BSM models involve extended scalar sectors with additional singlets, doublets, or higher multiplets. Some of these extensions introduce numerous new parameters that are only weakly constrained by experiment. Theoretical consistency therefore becomes essential.

An important illustration of these theoretical constraints is the requirement of the unitarity of the scattering matrix or S-matrix, which leads to the so-called perturbative unitarity bounds. The foundational work of Lee, Quigg, and Thacker~\cite{Lee:1977eg} demonstrated its power by deriving an upper bound on the SM Higgs mass. Their analysis leveraged the Goldstone boson equivalence theorem, which states that at high energies, the scattering amplitudes of longitudinal gauge bosons are equivalent to their corresponding Goldstone bosons. This allows unitarity constraints to be efficiently applied to the scalar sector of the theory. Note that perturbative unitarity constraints on generic Yukawa and vector interactions have been analysed in~\cite{Allwicher:2021rtd, Barducci:2023lqx}. Regarding the scalar sector, the conventional approach, widely used in studies of BSM models, is to consider the high-energy limit where the center-of-mass energy far exceeds all particle masses. In this regime, the calculation simplifies as diagrams with propagators are suppressed, leaving only contact interactions.

In this letter we present a general method, which systematically derives the $2\rightarrow 2$ scattering matrix in the high-energy limit for models with arbitrary gauge groups and field representations. We can then compute the eigenvalues of the S-matrix and apply the perturbative unitarity bounds on them. This method is given in our Mathematica package \texttt{anyPUB}. This package, accompanied with an example file, can be found at \url{https://github.com/nicobenincasa/anyPUB}. We then apply our framework to the minimal left-right symmetric model (MLRSM), deriving explicit constraints and comparing our results with the existing literature, showing that the previously obtained results are wrong. To further demonstrate the generality of our formalism, we employ it in the context of the Pati-Salam model, providing the first complete calculation of its perturbative unitarity constraints.

\section{Perturbative unitarity}
\label{sec:PU}

The partial-wave decomposition of the invariant amplitude for $2\rightarrow2$ scattering processes is given by
\begin{equation}
    \mathcal{M}(s,\theta)=16\pi\sum_{j=0}^\infty a_J(s)(2J+1)P_J(\cos\theta),
\end{equation}
with $\sqrt{s}$, the center-of-mass energy, $J$ the total orbital angular momentum of the final state and $P_J$ the Legendre polynomial of degree $J$.

Focusing on the zeroth partial (or $s$-)wave amplitude $a_0$, tree-level partial-wave unitarity imposes
\begin{equation}
    \left\vert\Ree(a_0^i)\right\vert\leq \frac{1}{2},\quad\text{with}\quad \mathcal{M}=16\pi a_0
\end{equation}
for each eigenvalue $a_0^i$. Therefore the task consists in computing the eigenvalues of $\mathcal{M}$ and constraining them as 
\begin{equation}
   \label{eq:PU}
   \boxed{\left\vert\Ree(\mathcal{M}_i)\right\vert\leq 8\pi}
\end{equation}

Thanks to the Goldstone boson equivalence theorem, one can easily derive the perturbative unitarity bounds in the limit of high energy, $s\rightarrow \infty$, via four-point vertices only, thus constraining the quartic couplings of the model. In that case, the S-matrix elements $\mathcal{M}_{AB,CD}\equiv \langle \phi_A\phi_B|\mathcal{M}|\phi_C\phi_D\rangle$, with $\phi_X$, $X=A,B,C,D$, a scalar field, can straightforwardly be computed by taking the fourth derivative of $V_4$, the latter containing only the quartic terms of the scalar part of the tree-level potential:
\begin{equation}
\label{eq:matrix_element}
    \mathcal{M}_{AB,CD}=\frac{1}{2^{(\delta_{AB}+\delta_{CD})/2}}\frac{-\partial^4V_4}{\partial\phi_A\partial\phi_B\partial\phi_C^*\partial\phi_D^*},
\end{equation}
with $\delta_{ij}$ the Kronecker delta and where a symmetry factor $1/\sqrt{2}$ is applied for initial/final states made of identical particles.

\section{Derivation of the block-diagonal S-matrix}
\label{sec:s-matrix}

The function \texttt{ScatteringMatrix} of our Mathematica package \texttt{anyPUB} allows to efficiently derive the $2\rightarrow 2$ scattering matrix in the basis made of real components of the fields for any models! One simply needs to provide the quartic potential $V_4$ expressed in terms of real scalar fields only, and indicate which quartic couplings are real. Then \texttt{anyPUB} simply extracts all the real fields $\phi_X$, $X=A,B,...$, constructs all possible unordered pairs $\phi_{XY}\equiv\phi_X\phi_Y$ and combines them to obtain an unordered set of $\phi_{XYVW}\equiv\phi_{XY}\phi_{VW}$. Next, it computes the derivative of $V_4$ with respect to each $\phi_{XYVW}$ in order to build the upper triangle of the S-matrix. It then generates the lower triangle, which is simply the conjugate transpose of the upper one and appropriately rescales the resulting full S-matrix with symmetry factors. Finally, this matrix is block-diagonalised by permuting columns and rows, using graph theory, via \texttt{MakeBlockDiagonal}. Ultimately, instead of diagonalising the full S-matrix, the eigenvalues can be extracted from each irreducible block either analytically or numerically, depending on the dimension and the sparsity of the block in question. Examples of how to use this package are given in Section~\ref{sec:package}.

\section{Discussion}
\label{sec:discussion}

We claim that \texttt{anyPUB} works for any models. We tested  it for a range of models such as SM + real/complex singlet~\cite{Kang:2013zba, Alanne:2020jwx}, $\mathbb{Z}_2$-symmetric 2HDM~\cite{Ginzburg:2003fe}, 2HD+a model~\cite{Arcadi:2022lpp}, IDM~\cite{Benincasa:2022elt}, dark $SU(2)$ model~\cite{Benincasa:2025tdr}, $\mathbb{Z}_3$-2HDM + complex singlet~\cite{Benincasa:2023vyp}, $U(1)\times U(1)$-symmetric 3HDM~\cite{Bento:2022vsb}, Higgs triplet model~\cite{Khan:2016sxm}, economical 331 model~\cite{Kannike:2025qru}, Georgi-Machacek model~\cite{Aoki:2007ah} and type-II seesaw model~\cite{Arhrib:2011uy},  where we could successfully obtain the analyical results found in the literature. However for models like general 2HDM~\cite{Ginzburg:2005dt} or $\mathbb{Z}_3$-symmetric 3HDM~\cite{Bento:2022vsb}, the size of some matrix blocks were too large to compute the corresponding eigenvalues analytically.

Actually for $SU(2)\otimes U(1)$ gauge theories with an arbitrary number of scalar fields, either singlets or doublets of $SU(2)$, a recent Mathematica package \texttt{BounDS} can systematically obtain a basis in terms of quantum numbers $\vert Q,Y,T\rangle$, thereby blocks of the scattering matrix are naturally smaller and it is then easier to extract the eigenvalues analytically~\cite{Lopes:2025krb}. While our method also works for such models, it becomes trickier to obtain all eigenvalues analytically for more complex versions of these models like the general NHDM. However, we can still use our method to obtain the eigenvalues numerically. Moreover, unlike \texttt{BounDS}, it also works for models involving any gauge groups and multiplet representations, e.g the minimal left-right symmetric model consisting in bi-doublets and triplets (see Section~\ref{sec:MLRSM}). 

Alternatively, one can use the \texttt{SARAH} Mathematica package~\cite{Goodsell:2018tti}, which is designed to compute a variety of theoretical quantities. In addition to perturbative unitarity analysis, SARAH can produce outputs such as mass matrices, renormalization group equations, and model files for external tools, enabling further phenomenological studies in combination with dedicated software. While SARAH is quite efficient when applied to pre-existing models, the construction and implementation of a new model demand substantial effort. Next, regarding perturbative unitarity, SARAH considers quantum numbers and conservation laws such that only states that are allowed to mix are considered when deriving the $2\rightarrow 2$ S-matrix. However, for models with large representations and many degrees of freedom, the process is computationally intensive and, in practice, unfeasible. Finally, note that the \texttt{anyPUB} function \texttt{MakeBlockDiagonal} that block-diagonalises matrices can also be used for large scattering matrices obtained with \texttt{SARAH}, in order to compute the eigenvalues more easily. Let us now illustrate our method with two specific models: the MLRSM and the Pati-Salam model.

\subsection{Minimal left-right symmetric model}
\label{sec:MLRSM}

The quartic part of the scalar potential of the minimal left-right symmetric model invariant under $SU(3)_C \otimes SU(2)_L \otimes SU(2)_R \otimes U(1)_{B-L}$ is given by~\cite{Deshpande:1990ip}
\begin{align}
    V_4 &= \lambda_1 \Tr^2\left[\Phi^\dagger\Phi\right] + \lambda_2 \left(\Tr^2\left[\tilde{\Phi}\Phi ^\dagger\right]+\Tr^2\left[\tilde{\Phi}^\dagger\Phi\right]\right) \nonumber \\
    &+ \lambda_3\Tr\left[\tilde{\Phi}\Phi^\dagger\right] \Tr\left[\tilde{\Phi}^\dagger\Phi\right]\nonumber \\
    & + \lambda_4\left\{\Tr\left[\Phi^\dagger\Phi\right]\left(\Tr\left[\tilde{\Phi}\Phi^\dagger\right] + \Tr\left[\tilde{\Phi}^\dagger\Phi\right]\right)\right\}\nonumber \\
    &+ \lambda_5\left(\Tr^2\left[\Delta_L\Delta_L^\dagger\right] + \Tr^2\left[\Delta_R\Delta_R^\dagger\right]\right)  \nonumber\\
    &+ \lambda_6\left(\Tr[\Delta_L\Delta_L] \Tr\left[\Delta_L^\dagger\Delta_L^\dagger\right] + \Tr[\Delta_R\Delta_R]\Tr\left[\Delta_R^\dagger\Delta_R^\dagger\right]\right)  \nonumber\\
    &+\lambda_7\Tr\left[\Delta_L\Delta_L^\dagger\right]\Tr\left[\Delta_R\Delta_R^\dagger\right] \nonumber \\
    & +\lambda_8\left(\Tr[\Delta_R\Delta_R]\Tr\left[\Delta_L^\dagger\Delta_L^\dagger\right] + \Tr[\Delta_L\Delta_L]\Tr\left[\Delta_R^\dagger\Delta_R^\dagger\right]\right) \nonumber \\
    &+ \lambda_9\Tr\left[\Phi^\dagger\Phi\right]\left(\Tr\left[\Delta_L^\dagger\Delta_L\right]+\Tr\left[\Delta_R^\dagger\Delta_R\right]\right)\nonumber\\
    &+\bigg\{(\lambda_{10}+i~\lambda_{11})\Big(\Tr\left[\Delta_L\Delta_L^\dagger\right]\Tr\left[\Phi ^\dagger\tilde{\Phi}\right]\nonumber\\
    &+\Tr\left[\Delta_R\Delta_R^\dagger\right] \Tr\left[\tilde{\Phi}^\dagger\Phi\right]\Big)+ \text{h.c.}\bigg\}\nonumber\\
    &+\lambda_{12}\left(\Tr\left[\Phi\Phi^\dagger\Delta_L\Delta_L^\dagger\right]+\Tr\left[\Phi ^\dagger\Phi\Delta_R\Delta_R^\dagger\right]\right),
\end{align}
where all the quartic couplings are real and where the fields are parametrised as
\begin{align}
    &\Phi=\begin{pmatrix}
    \phi_1^0 & \phi_1^+\\
    \phi_2^- & \phi_2^0 \end{pmatrix},\quad\tilde{\Phi}=\sigma_2\Phi^*\sigma_2,\nonumber\\
    &\Delta_{L,R} = \begin{pmatrix}
    \delta^+_{L,R}/\sqrt{2} & \delta^{++}_{L,R}\\
    \delta^0_{L,R} & -\delta^+_{L,R}/\sqrt{2} \end{pmatrix},
\end{align}
with quantum numbers given in Table~\ref{tab:fields1}.

\begin{table}[!t]
    \centering
    \begin{tabular}{c c c c c}
      Fields & SU(3)$_C$ & SU(2)$_L$ & SU(2)$_R$ & U(1)$_{B-L}$  \\ \hline
        $\Phi$ & \textbf{1} & \textbf{2} & \textbf{2} & 0 \\
        $\Delta_L$ & \textbf{1} & \textbf{3} & \textbf{1} & 2 \\
        $\Delta_R$ & \textbf{1} & \textbf{1} & \textbf{3} & 2 
    \end{tabular}
    \caption{Scalar fields and their quantum numbers in the minimal left-right symmetric model.}
    \label{tab:fields1}
\end{table}

Perturbative unitarity constraints have been derived in~\cite{Mondal:2015fja}. Let us now check if we obtain the same results using our method. We obtain a $210\times 210$ block-diagonal S-matrix via the function \texttt{ScatteringMatrix}. It is made of 22 blocks $B_{n}$ of dimensions $n\times n$: one $B_{36}$, four $B_{24}$, two $B_{16}$, one $B_{6}$, eight $B_{4}$, two $B_{2}$ and four $B_{1}$. By a random numerical substitution, we can directly check that only 21 distinct eigenvalues are associated to this $210\times 210$ block-diagonal S-matrix. We then compute the eigenvalues of the four $B_{24}$ using \texttt{MatrixMinimalPolynomial}, which returns the same minimal polynomial of degree 4 for each and we then solve the latter. Combining with the smaller blocks $B_i$, $i\leq 16$, we find 16 distinct eigenvalues in total. We can then obtain the 5 remaining eigenvalues from the last block $B_{36}$  as follows. We first identify numerically which of the 16 eigenvalues are eigenvalues of $B_{36}$ as well as their multiplicity. Next, since the trace of the power of a matrix $B$ yields the relation
\begin{equation}
    \Tr ~B^k = \sum_{i=1}^N m_i~\Lambda_i^k,
\end{equation}
with $N$ the number of distinct eigenvalues and $m_i$ the multiplicity of the eigenvalue $\Lambda_i$ of the matrix $B$, we can then define the symmetric polynomial
\begin{equation}
    p_k \equiv \Tr ~B^k - \sum_{i=1}^M m_i~\Lambda_i^k = \sum_{i=1}^{N-M}x_i^k,
\end{equation}
with $M$ the number of distinct known eigenvalues and $x_i$ the unknown eigenvalues.
However, instead of solving the system of $N-M$ equations $p_k-\sum_{i=1}^{N-M}x_i^k=0$ for $k=1,\dots,N-M$, we can simplify the problem using Newton's identities. They relate the symmetric polynomial $p_k=\sum_{i=1}^{N-M}x_i^k$ to the elementary symmetric polynomial $e_k$, with $e_0=1$, as follows~\cite{seroul2000newton}:
\begin{equation}
    e_k=\frac{1}{k}\sum_{i=1}^k(-1)^{i-1}e_{k-i}~p_i.
\end{equation}
After computing $e_1,\dots,e_{N-M}$, we simply need to solve the equation
\begin{equation}
    R(x)=\prod_i^{N-M}(x-x_i)=\sum_{k=0}^{N-M}(-1)^k ~e_k ~x^{{N-M}-k}=0,
\end{equation}
in order to derive the $N-M$ remaining eigenvalues. 

Applying this method, with \texttt{ReducedPolynomialNew-\\ton}, to the matrix $B_{36}$, with $N-M=5$, we can easily compute the 5 remaining eigenvalues. We find that two of them have a simple analytical expression. We thus compute the polynomial quotient to obtain the polynomial of degree 3 in Eq.(\ref{eq:cubic_eq}), via \texttt{ReducedCharacteristicPolynomial}. The 21 distinct eigenvalues are given by
\begin{align}
\label{eq:MLRSM_eig1}
  \Lambda_{1,2} &=   \lambda_{9}+\frac{\lambda_{12}}{2}\pm\sqrt{4 \left(\lambda_{10}^2+\lambda_{11}^2\right)+\lambda_{12}^2},\\
  \Lambda_{3,4} &= \lambda_{9}+\frac{\lambda_{12}}{2} \pm\sqrt{4 \left(\lambda_{10}^2+\lambda_{11}^2\right)+\frac{\lambda_{12}^2}{2}},\\
  \Lambda_{5,6} &= 2 \bigg(\lambda_{1}+2 \lambda_{2}+\lambda_{3}\pm\sqrt{(\lambda_{3}-2 \lambda_{2})^2+4 \lambda_{4}^2}\bigg),\\
  \Lambda_{7,8} &= \lambda_{1}-2 \lambda_{3}+\lambda_{5}-2 \lambda_{6}\nonumber\\
  &~~~~\pm\sqrt{(\lambda_{1}-2 \lambda_{3}-\lambda_{5}+2 \lambda_{6})^2+2 \lambda_{12}^2},\\ 
  \Lambda_{9,10} &=  \lambda_{1}-12 \lambda_{2}+4 \lambda_{3}+4 \lambda_{5}+2 \lambda_{6}-\frac{3}{2} \lambda_{7}\nonumber\\
  &~~~~\pm \frac{1}{2}\big[(2 \lambda_{1}-4 (6 \lambda_{2}-2 \lambda_{3}+2 \lambda_{5}+\lambda_{6})+3 \lambda_{7})^2\nonumber\\
  &~~~~~~~~~~~+384 \lambda_{11}^2\big]^{1/2},\\
  \Lambda_{11,12} &= 2 \left[\lambda_{5}+3 (\lambda_{6}\pm\lambda_{8})\right],\\
  \Lambda_{13,14} &= \lambda_{7}\pm 4 \lambda_{8},\\
  \Lambda_{15} &= 2 (\lambda_{1}-4 \lambda_{2}),\\
  \Lambda_{16} &= 2 (\lambda_{5}+2 \lambda_{6}),\\
  \Lambda_{17} &= 2\lambda_5,\\
  \label{eq:MLRSM_eig18}
  \Lambda_{18} &= \lambda_7,
\end{align}
and the three solutions of the cubic equation
\begin{align}
\label{eq:cubic_eq}
    0 =&~ x^3 + \left(-12 \lambda_{1}-24 \lambda_{2}-12 \lambda_{3}-8 \lambda_{5}-4 \lambda_{6}-3 \lambda_{7}\right)x^2 \nonumber \\
    &+(20 \lambda_{1}^2+240 \lambda_{1} \lambda_{2}+88 \lambda_{1} \lambda_{3}+96 \lambda_{1} \lambda_{5}+48 \lambda_{1} \lambda_{6}\nonumber \\
    &+36 \lambda_{1} \lambda_{7}-96 \lambda_{10}^2-6 \lambda_{12}^2-24 \lambda_{12} \lambda_{9}+96 \lambda_{2} \lambda_{3}\nonumber\\
    &+192 \lambda_{2} \lambda_{5}+96 \lambda_{2} \lambda_{6}+72 \lambda_{2} \lambda_{7}+32 \lambda_{3}^2+96 \lambda_{3} \lambda_{5}\nonumber \\
    &+48 \lambda_{3} \lambda_{6}+36 \lambda_{3} \lambda_{7}-144 \lambda_{4}^2-24 \lambda_{9}^2) x \nonumber\\
    &-160 \lambda_{1}^2 \lambda_{5}-80 \lambda_{1}^2 \lambda_{6}-60 \lambda_{1}^2 \lambda_{7}+960 \lambda_{1} \lambda_{10}^2\nonumber \\
    &+12 \lambda_{1} \lambda_{12}^2+48 \lambda_{1} \lambda_{12} \lambda_{9}-1920 \lambda_{1} \lambda_{2} \lambda_{5}\nonumber\\
    &-960 \lambda_{1} \lambda_{2} \lambda_{6}-720 \lambda_{1} \lambda_{2} \lambda_{7}-704 \lambda_{1} \lambda_{3} \lambda_{5}\nonumber \\
    &-352 \lambda_{1} \lambda_{3} \lambda_{6}-264 \lambda_{1} \lambda_{3} \lambda_{7}+48 \lambda_{1} \lambda_{9}^2+384 \lambda_{10}^2 \lambda_{3}\nonumber\\
    &-576 \lambda_{10} \lambda_{12} \lambda_{4}-1152 \lambda_{10} \lambda_{4} \lambda_{9}+144 \lambda_{12}^2 \lambda_{2}\nonumber \\
    &+48 \lambda_{12}^2 \lambda_{3}+576 \lambda_{12} \lambda_{2} \lambda_{9}+192 \lambda_{12} \lambda_{3} \lambda_{9}\nonumber\\
    &-768 \lambda_{2} \lambda_{3} \lambda_{5}-384 \lambda_{2} \lambda_{3} \lambda_{6}-288 \lambda_{2} \lambda_{3} \lambda_{7}\nonumber \\
    &+576 \lambda_{2} \lambda_{9}^2-256 \lambda_{3}^2 \lambda_{5}-128 \lambda_{3}^2 \lambda_{6}-96 \lambda_{3}^2 \lambda_{7}\nonumber\\
    &+192 \lambda_{3} \lambda_{9}^2+1152 \lambda_{4}^2 \lambda_{5}+576 \lambda_{4}^2 \lambda_{6}+432 \lambda_{4}^2 \lambda_{7},
\end{align}
to which we impose the perturbative unitarity constraint~(\ref{eq:PU}): $\vert \Lambda_i\vert\leq 8\pi,\forall i$, where we omit $\Ree$ because $\mathcal{M}$ is hermitian and thus has real eigenvalues.

The form of these eigenvalues is much simpler than the one found in~\cite{Mondal:2015fja}. Moreover, by classifying states by electric charge, they found 64 independent constraints for perturbative unitarity, while our method provides 21 distinct eigenvalues only! We find only 5 similar eigenvalues among the 64 ones:
\begin{equation}
    2 \lambda_5,\quad 2 (\lambda_5 + 2 \lambda_6),\quad \lambda_7,\quad \lambda_7 \pm 4 \lambda_8.
\end{equation}

Next we analyse the Mathematica notebook provided in their paper~\cite{Mondal:2015fja}. The source of discrepancies actually comes from the fact that in their potential they define $\tilde{\Phi}=\sigma_2\Phi\sigma_2$ instead of $\tilde{\Phi}=\sigma_2\Phi^*\sigma_2$. Moreover, their $\lambda_9$ term contains $\Tr\left[\Delta_L^\dagger\Delta_L^\dagger\right]+\Tr\left[\Delta_R^\dagger\Delta_R^\dagger\right]$, instead of $\Tr\left[\Delta_L^\dagger\Delta_L\right]+\Tr\left[\Delta_R^\dagger\Delta_R\right]$, while their $\lambda_{12}$ term contains $\Tr\left[\Phi^\dagger\Phi\Delta_L\Delta_L^\dagger\right]$ instead of $\Tr\left[\Phi\Phi^\dagger\Delta_L\Delta_L^\dagger\right]$. Additionally, we find that in their implementation, the symmetry factor present in Eq.~(\ref{eq:matrix_element}) is not taken into account. Furthermore, in their classification of single-charged states they are mixing both (+1)-charged states and (-1)-charged states. We have checked that, using their code with all these modifications applied, we indeed recover our 21 distinct eigenvalues displayed in Eq.~(\ref{eq:MLRSM_eig1})-(\ref{eq:MLRSM_eig18}) and Eq.~(\ref{eq:cubic_eq}).

\begin{figure}[!h]
    \centering
    \includegraphics[width=0.8\linewidth]{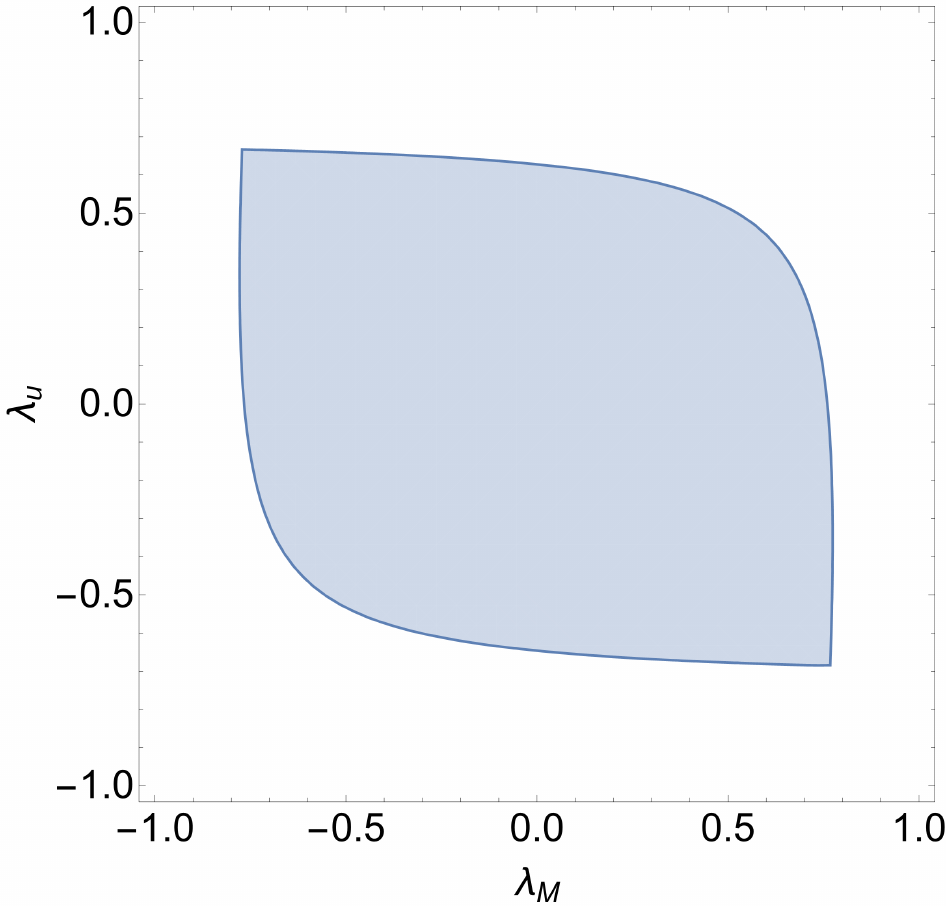}
    \caption{Region of the parameter space allowed by perturbative unitarity bounds in the MLRSM.}
    \label{fig:pub_plot1}
\end{figure}

Finally, let us show how these perturbative unitarity bounds constrain the parameter space in the MLRSM. In this model, the mass of the heavy scalar states is made of one leading-order contribution proportional to the vacuum expectation value (VEV) $v_R$ from $\langle \Delta_R\rangle$, while the sub-leading contribution is proportional to the electroweak VEV $v_\text{EW}$, with $v_\text{EW}\ll v_R$. The $v_R$ part of these masses is proportional to a linear combination of $\lambda_5, \lambda_6,\lambda_7,\lambda_{12}$, while $\lambda_2, \lambda_3,\lambda_4,\lambda_8,\lambda_9,\lambda_{10},\lambda_{11}$ only contribute to the subdominant part proportional to $v_\text{EW}$~\cite{Mondal:2015fja} (note that $\lambda_{12}$ also contributes to it~\cite{Deshpande:1990ip}). Given that this model contains 12 quartic couplings, we assume a mass degeneracy at leading-order for the heavy scalars, thus implying $\lambda_M = \lambda_5 = \lambda_6 = (\lambda_7-2\lambda_5)/4=\lambda_{12}/4$. Following~\cite{Mondal:2015fja}, we also consider $\lambda_u = \lambda_2=\lambda_3=\lambda_4=\lambda_8=\lambda_9=\lambda_{10}=\lambda_{11}$. In Figure~\ref{fig:pub_plot1}, we show the region of the parameter space, in the plane $\lambda_u-\lambda_M$, which satisfies the perturbative unitarity bounds in the MLRSM. We can see that these bounds constrain the quartic couplings to be small: $\vert\lambda_u\vert<0.7$ and $\vert\lambda_M\vert<0.8$, thus implying $M\lesssim 1.25 ~v_R$, with $M$ the mass of the heavy scalar states.

\subsection{Pati-Salam model}
\label{sec:Pati-Salam}

The quartic part of the scalar potential of the Pati-Salam model invariant under $SU(4) \otimes SU(2)_L\otimes SU(2)_R$ is given by~\cite{Giudice:2014tma}
\begin{align}
V_4=&~\lambda_1\Tr^2\left[\Phi^\dagger\Phi\right]+\Ree\left(\lambda_2\Tr^2\left[\Phi^\dagger\Phi^c\right]\right)\nonumber\\
    &+\Ree\left(\lambda_3\Tr\left[\Phi^\dagger\Phi\right]\Tr\left[\Phi^\dagger\Phi^c\right]\right)\nonumber\\
    &+(\lambda_4-2\Ree\lambda_2)\left\vert\Tr\left[\Phi^\dagger\Phi^c\right]\right\vert^2 \nonumber\\
    &+\lambda_{R1}\Tr^2\left[\phi_R^\dagger\phi_R\right]+\lambda_{R2}\Tr\left[\phi_R^\dagger\phi_R\phi_R^\dagger\phi_R\right]\nonumber\\
    &+\lambda_{R\Phi 1} \Tr\left[\phi_R^\dagger\phi_R\right]\Tr\left[\Phi^\dagger\Phi\right] \nonumber\\
    &+\Ree\left(\lambda_{R\Phi 2}\Tr\left[\phi_R^\dagger\phi_R\right]\Tr\left[\Phi^\dagger\Phi^c\right]\right)\nonumber\\
    &+\lambda_{R\Phi 3}\Tr\left[\phi_R\phi_R^\dagger\Phi^\dagger\Phi\right],
\end{align}
where quantum numbers of the fields are given in Table~\ref{tab:fields2}.

\begin{table}[!t]
    \centering
    \begin{tabular}{c c c c}
      Fields & SU(4) & SU(2)$_L$ & SU(2)$_R$   \\ \hline
        $\phi_R$ & \textbf{4} & \textbf{1} & \textbf{2} \\
        $\Phi$ & \textbf{1} & \textbf{2} & \textbf{2} 
    \end{tabular}
    \caption{Scalar fields and their quantum numbers in the Pati-Salam model.}
    \label{tab:fields2}
\end{table}

To our knowledge, perturbative unitarity bounds have not yet been derived within the frame of the Pati-Salam model. Our function \texttt{ScatteringMatrix} yields a $300\times 300$ block-diagonal S-matrix. It is made of 42 blocks $B_{n}$ of dimensions $n\times n$: one $B_{36}$, one $B_{32}$, eight $B_{16}$, twenty-four $B_{4}$ and eight $B_{1}$. As for the MLRSM in Section~\ref{sec:MLRSM}, we can numerically check that this $300\times 300$ matrix only has 23 distinct eigenvalues. We obtain 10 distinct eigenvalues from the blocks $B_{16}, B_{4}$ and $B_{1}$:
\begin{align}
  \Lambda_{1,2} &=   \frac{1}{2} \left(2 \lambda_{R\phi 1}+\lambda_{R\phi 3}\pm\sqrt{4 \vert\lambda_{R\phi 2}\vert^2+\lambda_{R\phi 3}^2}\right),\\
  \Lambda_{3,4} &= \frac{1}{2} \left(2 \lambda_{R\phi 1}+\lambda_{R\phi 3}\pm\sqrt{4 \vert\lambda_{R\phi 2}\vert^2+9\lambda_{R\phi 3}^2}\right),\\
  \Lambda_{5,6} &=2 (\lambda_{R1}\pm\lambda_{R2}),\\
  \Lambda_7 &=2 (\lambda_{R1}+2\lambda_{R2}),\\
  \Lambda_8 &= \lambda_{R\phi 1}+\lambda_{R\phi 3},\\
  \Lambda_9 &= 2\lambda_{R1},\\
  \Lambda_{10} &= \lambda_{R\phi1}
  \end{align}

The remaining 13 distinct eigenvalues contained in $B_{36}$ and $B_{32}$ should be computed numerically.

\begin{figure}[!h]
    \centering
    \includegraphics[width=0.8\linewidth]{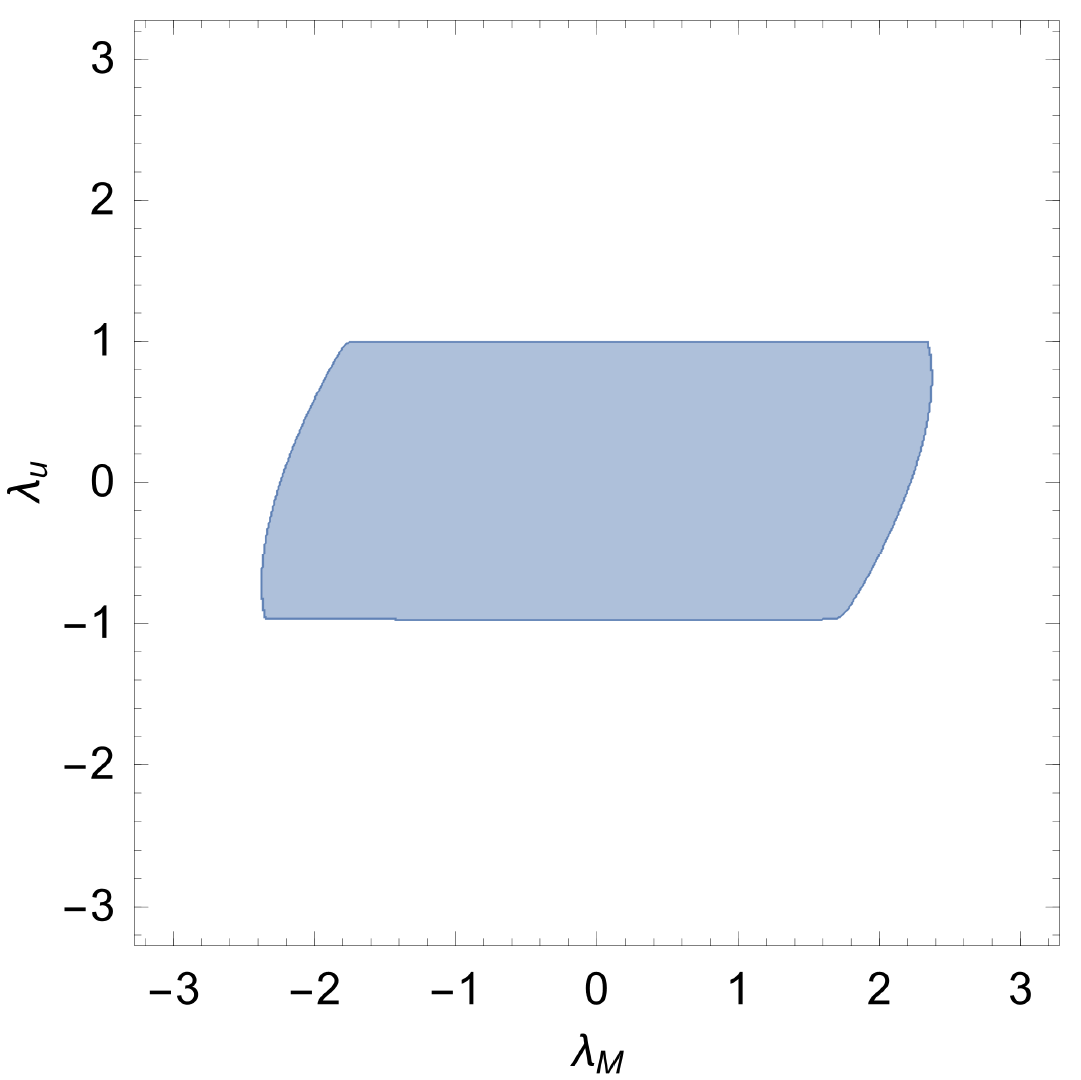}
    \caption{Region of the parameter space allowed by perturbative unitarity bounds in the Pati-Salam model.}
    \label{fig:pub_plot2}
\end{figure}

Finally, let us show how these perturbative unitarity bounds constrain the parameter space in the Pati-Salam model. In this model, the mass of the heavy scalar states is made of one leading-order contribution proportional to the vacuum expectation value  $v_R$ from $\langle \phi_R\rangle$, while the sub-leading contribution is proportional to the electroweak VEV $v_\text{EW}$, with $v_\text{EW}\ll v_R$. The $v_R$ part of these masses is proportional to a linear combination of $\lambda_{R1}, \lambda_{R2},\lambda_{R\phi3}$, while $\lambda_2, \lambda_3,\lambda_4,\lambda_{R\phi1},\lambda_{R\phi2}$ only contribute to the subdominant part proportional to $v_\text{EW}$. As in the previous section we then assume a mass degeneracy at leading-order for the heavy scalars, thus implying $\lambda_M = \lambda_{R1} = -3\lambda_{R2}/2 = -(\lambda_{R\phi3}+4\lambda_{R2})/4$. Next we also consider $\lambda_u = \lambda_2=\lambda_3=\lambda_4=\lambda_{R\phi1}=\lambda_{R\phi2}$. In Figure~\ref{fig:pub_plot2}, we show the region of the parameter space, in the plane $\lambda_u-\lambda_M$, which satisfies the perturbative unitarity bounds in the Pati-Salam model. We can see that these bounds constrain the quartic couplings to be small, although the bounds are looser than in the MLRSM: $\vert\lambda_u\vert<1$ and $\vert\lambda_M\vert<2.4$, thus implying $M\lesssim 2.2 ~v_R$, with $M$ the mass of the heavy scalar states.

\section{Conclusions}
\label{sec:conclusion}

In this letter, we have provided a package \texttt{anyPUB} which systematically derives the $2\rightarrow 2$ scattering matrix in the high-energy limit, via the function \texttt{ScatteringMatrix}, for any models, irrespective of the gauge group or the multiplet representation. Then, \texttt{MakeBlockDiagonal} makes this matrix block-diagonal, so that we only need to treat each irreducible block individually when computing the eigenvalues. The list of derived eigenvalues is obtained with \texttt{EigenvaluesScatteringMatrix}. If some matrix blocks have a too large dimension or are not sparse enough, then we should opt for an alternative way to derive the eigenvalues, among  \texttt{MatrixMinimalPolynomial}, \texttt{ReducedCharacteristicPolynomial} and \texttt{ReducedPol-\\ynomialNewton}. In the case where none of these methods are conclusive, then we can still obtain the eigenvalues numerically. 

We have derived the perturbative unitarity bounds in the case of the minimal left-right symmetric model and shown that our results are very different than those of~\cite{Mondal:2015fja}, highlighting the sources of discrepancies. We obtained only 21 independent eigenvalues compared to 64 in their paper. Moreover, the expressions of our eigenvalues are much simpler than theirs. Next, we derived for the first time the perturbative unitarity bounds in the Pati-Salam model. We could obtain 10 eigenvalues analytically, while the 13 remaining ones should be computed numerically. Finally, for both models, we have graphically shown how these perturbative unitarity bounds constrain the parameter space.

We believe that \texttt{anyPUB} can be easily applied to any other BSM models and hope that it will be useful to the community.

\section*{Acknowledgments}
We deeply thank Kristjan Kannike and Kristjan Müürsepp for helpful discussions and for reading the manuscript. NB is supported by the National Natural Science Foundation of China (Grant No. 12475105).

\appendix

\section{How to use \texttt{anyPUB}?}
\label{sec:package}

In this section, we provide a detailed explanation of how to obtain the eigenvalues of the S-matrix, in two specific models.

\subsection{Definition of functions}
In this section, we provide a description for the functions used in \texttt{anyPUB}: 

\begin{itemize}
  \item \texttt{ScatteringMatrix[quarticPotential, assump- tion]}, with \texttt{quarticPotential} the quartic part of the scalar potential and \texttt{assumption} specifying which quartic couplings are real, returns an association between keys and values: \texttt{FieldList} is the list of independent real scalar fields in \texttt{quarticPotential}, \texttt{PairsList} is the list of all unordered pairs of these fields, \texttt{ScatteringPairs} is a nested list of all possible unordered combinations of these pairs, \texttt{Vertices} is a nested list computed four-point vertices for each scattering pair and \texttt{ScatteringMatrix} is the S-matrix.

  \item \texttt{MakeBlockDiagonal[matrix]}, with \texttt{matrix} the matrix to diagonalise, returns an association between keys and values: \texttt{AdjacencyMatrix} is the adjacency matrix obtained from \texttt{matrix}, \texttt{Graph} is the graph associated to this adjacency matrix, \texttt{Components} is the list of rows/columns index used for each connected subgraphs of the graph and \texttt{BlockMatrix} is the block-diagonalised version of \texttt{matrix}.

  \item \texttt{EigenvaluesScatteringMatrix[quarticPotent- ial, assumption]}, with \texttt{quarticPotential} the quartic part of the scalar potential and \texttt{assumption} specifying which quartic couplings are real, returns an association between keys and values: \texttt{EigenvaluesPerBlock} is the list of eigenvalues per irreducible block of the block-diagonalised matrix, while \texttt{AllDistinctEigenvalues} provide the list of distinct eigenvalues of the S-matrix.

  \item \texttt{MatrixMinimalPolynomial[matrix, x, number- DistinctEigenvalues]} returns the minimal polynomial of \texttt{matrix} using \texttt{x} as a variable. The number of distinct eigenvalues,  \texttt{number- DistinctEigenvalues} is optional, it is set to 0 by default. If it is larger than 0, then this function avoids computing the null space of \texttt{matrix} as long as the number of iterations is smaller than or equal to  \texttt{numberDistinctEigenvalues}.

  \item \texttt{ReducedCharacteristicPolynomial[character- isticPolynomial, eigenvaluesList,~x]} returns the polynomial quotient of \texttt{character- isticPolynomial} in terms of \texttt{x}, by using a set of known eigenvalues, \texttt{eigenvaluesList}.

  \item \texttt{ReducedPolynomialNewton[matrix, eigenvalu- esList, multiplicityList, x]} returns a polynomial in terms of \texttt{x}, via the Newton's identities, by using a set a known eigenvalues, \texttt{eigenvaluesList}, and their corresponding multiplicity, \texttt{multiplicityList}.

\end{itemize}

\subsection{Example 1: Standard Model +  real singlet with $\mathbb{Z}_2$ symmetry}

Let us first try with the easiest extension of the Standard Model, by adding a real scalar singlet to the SM. The $\mathbb{Z}_2$-invariant quartic potential is given by
\begin{align}
    &V_4 = \frac{\lambda_H}{2}\left(H^\dagger H\right)^2+\frac{\lambda_S}{2}S^4+\frac{\lambda_{HS}}{2}\left(H^\dagger H\right)S^2,\nonumber\\
    &\text{with}\quad H=\frac{1}{\sqrt{2}}\begin{pmatrix}
    \omega_1+i~\omega_2 \\
    h+i~z\end{pmatrix}
\end{align}
 denoting the SM Higgs doublet and $S$ a real scalar singlet. All quartic couplings are taken to be real. We now have everything to derive the S-matrix, simple as that! Let us see now how to use \texttt{anyPUB}.\newline

Load the package:
\begin{lstlisting}[language=Mathematica]
<< anyPUB.wl
\end{lstlisting}

Express the quartic part of the tree-level potential in terms of real fields and specify which couplings are real:
\begin{lstlisting}[language=Mathematica]
V4 = \[Lambda]H/2 (H\[Conjugate].H)^2 + \[Lambda]S/2 S^4 +  \[Lambda]HS/2(H\[Conjugate].H) S^2;
VSinglet = ComplexExpand[V4 /. H -> 1/Sqrt[2] {\[Omega]1 + I \[Omega]2, h + I z}];
assumption = {{\[Lambda]H, \[Lambda]S, \[Lambda]HS} \[Element] Reals};
\end{lstlisting}

Derive the $2 \rightarrow 2$ scattering matrix in the $s\rightarrow\infty$ limit via the four-point vertices:
\begin{lstlisting}[language=Mathematica]
sm = ScatteringMatrix[VSinglet, assumption]["ScatteringMatrix"]
\end{lstlisting}

\begin{equation}
 \resizebox{.99\hsize}{!}{$\begin{pmatrix}
        \frac{3\lambda_H}{2} & 0 & 0 & 0 & 0 & \frac{\lambda_H}{2} & 0 & 0 & 0 & \frac{\lambda_{HS}}{2} & 0 & 0 & \frac{\lambda_H}{2} & 0 & \frac{\lambda_H}{2}\\
        0 & \lambda_H & 0 & 0 & 0 & 0 & 0 & 0 & 0 & 0 & 0 & 0 & 0 & 0 & 0 \\
        0 & 0 & \lambda_{HS} & 0 & 0 & 0 & 0 & 0 & 0 & 0 & 0 & 0 & 0 & 0 & 0 \\
        0 & 0 & 0 & \lambda_H & 0 & 0 & 0 & 0 & 0 & 0 & 0 & 0 & 0 & 0 & 0 \\
        0 & 0 & 0 & 0 & \lambda_H & 0 & 0 & 0 & 0 & 0 & 0 & 0 & 0 & 0 & 0 \\
        \frac{\lambda_H}{2} & 0 & 0 & 0 & 0 & \frac{3\lambda_H}{2} & 0 & 0 & 0 & \frac{\lambda_{HS}}{2}& 0 & 0 & \frac{\lambda_H}{2} & 0 & \frac{\lambda_H}{2} \\
        0 & 0 & 0 & 0 & 0 & 0 & \lambda_{HS} & 0 & 0 & 0 & 0 & 0 & 0 & 0 & 0 \\
        0 & 0 & 0 & 0 & 0 & 0 & 0 & \lambda_H & 0 & 0 & 0 & 0 & 0 & 0 & 0 \\
        0 & 0 & 0 & 0 & 0 & 0 & 0 & 0 & \lambda_H & 0 & 0 & 0 & 0 & 0 & 0 \\
        \frac{\lambda_{HS}}{2} & 0 & 0 & 0 & 0 & \frac{\lambda_{HS}}{2} & 0 & 0 & 0 & 6\lambda_S & 0 & 0 & \frac{\lambda_{HS}}{2} & 0 & \frac{\lambda_{HS}}{2} \\
        0 & 0 & 0 & 0 & 0 & 0 & 0 & 0 & 0 & 0 & \lambda_{HS} & 0 & 0 & 0 & 0 \\
        0 & 0 & 0 & 0 & 0 & 0 & 0 & 0 & 0 & 0 & 0 & \lambda_{HS} & 0 & 0 & 0 \\
        \frac{\lambda_H}{2} & 0 & 0 & 0 & 0 & \frac{\lambda_H}{2} & 0 & 0 & 0 & \frac{\lambda_{HS}}{2} & 0 & 0 & 6\lambda_H & 0 & \frac{\lambda_H}{2} \\
        0 & 0 & 0 & 0 & 0 & 0 & 0 & 0 & 0 & 0 & 0 & 0 & 0 & \lambda_H & 0 \\
        \frac{\lambda_H}{2} & 0 & 0 & 0 & 0 & \frac{\lambda_H}{2} & 0 & 0 & 0 & \frac{\lambda_{HS}}{2} & 0 & 0 & \frac{\lambda_H}{2} & 0 & 6\lambda_H 
    \end{pmatrix}$}
\end{equation}

In order to block diagonalise the scattering matrix, convert it into an adjacency matrix by putting all non-zero off-diagonal elements of the S-matrix to 1, and the remaining elements to zero~\cite{harary1959graph}. This adjacency matrix can be obtained via
\begin{lstlisting}[language=Mathematica]
smBlock = MakeBlockDiagonal[sm];
smBlock["AdjacencyMatrix"]
\end{lstlisting}

The adjacency matrix $A$ of a graph should be interpreted as the following: an non-zero element $A_{ij}$ corresponds to an oriented edge $i\rightarrow j$ in the graph, from the vertex labeled by $i$ to the vertex labeled by $j$. The scattering matrix is hermitian at tree-level, thus the resulting adjacency matrix is symmetric, leading to an undirected graph (for the vertex $i$ and $j$, one both has the edge $i \rightarrow j$ and the edge $j \rightarrow i$). Therefore we only focus on the upper triangle. The non zero-elements of the adjacency matrix are at the following positions:
\begin{table}[!h]
    \centering
\begin{tabular}{c c c c}
     (1, 6) & (1, 10) & (1, 13) & (1, 15) \\
     (6, 10) & (6, 13) & (6, 15) & \\ 
     (10, 13) & (10, 15) & &\\
     (13, 15) & & &
\end{tabular}
\end{table}

The corresponding graph has 5 vertices, all connected with each other. The columns/rows 2, 3, 4, 5, 7, 8, 9, 11, 12, 14 containing only zero elements thus simply correspond to isolated vertices in the graph. The corresponding graph is obtained via
\begin{lstlisting}[language=Mathematica]
smBlock["Graph"]
\end{lstlisting}

\begin{figure}[!h]
\centering
\includegraphics[scale=0.75,trim = 1.59cm 0cm 0cm 0cm, clip]{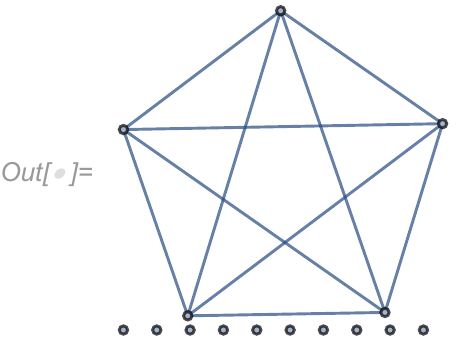}
\end{figure}

This graph consists of 11 components or connected subgraphs, 10 of which are just isolated vertices.
\begin{lstlisting}[language=Mathematica]
smBlock["Components"]
\end{lstlisting}
\begin{align}
      &  \{\{1, 6, 10, 13, 
  15\},\nonumber\\
  & \{14\}, \{12\}, \{11\}, \{9\}, \{8\}, \{7\}, \{5\}, \{4\}, \{3\}, \{2\}\}
\end{align}

By permuting rows/columns of the scattering matrix following the vertex labels of each component of the graph, we obtain the irreducible blocks of the scattering matrix:
\begin{lstlisting}[language=Mathematica]
MatrixForm /@ (sm[[#, #]] & /@ smBlock["Components"])
\end{lstlisting}
\begin{align}
   &\begin{pmatrix}
        \frac{3\lambda_H}{2} & \frac{\lambda_H}{2} & \frac{\lambda_{HS}}{2} &\frac{\lambda_H}{2} & \frac{\lambda_H}{2} \\ 
        \frac{\lambda_H}{2} & \frac{3\lambda_H}{2} & \frac{\lambda_{HS}}{2} &\frac{\lambda_H}{2} & \frac{\lambda_H}{2} \\ 
        \frac{\lambda_{HS}}{2} & \frac{\lambda_{HS}}{2} & 6\lambda_{S} &\frac{\lambda_{HS}}{2} & \frac{\lambda_{HS}}{2} \\ 
        \frac{\lambda_H}{2} & \frac{\lambda_H}{2} & \frac{\lambda_{HS}}{2} &\frac{3\lambda_H}{2} & \frac{\lambda_H}{2} \\ 
        \frac{\lambda_H}{2} & \frac{\lambda_H}{2} & \frac{\lambda_{HS}}{2} &\frac{\lambda_H}{2} & \frac{3\lambda_H}{2} 
        \end{pmatrix},(\lambda_H),~ (\lambda_{HS}),~ (\lambda_{HS}),\nonumber\\
        & (\lambda_{H}),(\lambda_{H}),~ (\lambda_{HS}),~ (\lambda_{H}),~ (\lambda_{H}),~ (\lambda_{HS}),~ (\lambda_{H})
\end{align}

The resulting block-diagonalised scattering matrix is given by
\begin{lstlisting}[language=Mathematica]
MatrixForm[smBlock["BlockMatrix"]]
\end{lstlisting}
\begin{equation}
   \resizebox{.99\hsize}{!}{$ \begin{pmatrix}
        \frac{3\lambda_H}{2} & \frac{\lambda_H}{2} & \frac{\lambda_{HS}}{2} &\frac{\lambda_H}{2} & \frac{\lambda_H}{2} & 1 & 0 & 0 & 0 & 0 & 0 & 0 & 0 & 0 & 0\\
        \frac{\lambda_H}{2} & \frac{3\lambda_H}{2} & \frac{\lambda_{HS}}{2} &\frac{\lambda_H}{2} & \frac{\lambda_H}{2} & 0 & 0 & 0 & 0 & 0 & 0 & 0 & 0 & 0 & 0 \\
        \frac{\lambda_{HS}}{2} & \frac{\lambda_{HS}}{2} & 6\lambda_{S} &\frac{\lambda_{HS}}{2} & \frac{\lambda_{HS}}{2} & 0 & 0 & 0 & 0 & 0 & 0 & 0 & 0 & 0 & 0 \\
        \frac{\lambda_H}{2} & \frac{\lambda_H}{2} & \frac{\lambda_{HS}}{2} &\frac{3\lambda_H}{2} & \frac{\lambda_H}{2}  & 0 & 0 & 0 & 0 & 0 & 0 & 0 & 0 & 0 & 0 \\
        \frac{\lambda_H}{2} & \frac{\lambda_H}{2} & \frac{\lambda_{HS}}{2} &\frac{\lambda_H}{2} & \frac{3\lambda_H}{2}  & 0 & 0 & 0 & 0 & 0 & 0 & 0 & 0 & 0 & 0 \\
        0 & 0 & 0 & 0 & 0 & \lambda_{H} & 0 & 0 & 0 & 0 & 0 & 0 & 0 & 0 & 0 \\
        0 & 0 & 0 & 0 & 0 & 0 & \lambda_{HS} & 0 & 0 & 0 & 0 & 0 & 0 & 0 & 0 \\
        0 & 0 & 0 & 0 & 0 & 0 & 0 & \lambda_{HS} & 0 & 0 & 0 & 0 & 0 & 0 & 0 \\
        0 & 0 & 0 & 0 & 0 & 0 & 0 & 0 & \lambda_{H} & 0 & 0 & 0 & 0 & 0 & 0 \\
        0 & 0 & 0 & 0 & 0 & 0 & 0 & 0 & 0 & \lambda_{H} & 0 & 0 & 0 & 0 & 0 \\
        0 & 0 & 0 & 0 & 0 & 0 & 0 & 0 & 0 & 0 & \lambda_{HS} & 0 & 0 & 0 & 0 \\
        0 & 0 & 0 & 0 & 0 & 0 & 0 & 0 & 0 & 0 & 0 & \lambda_{H} & 0 & 0 & 0 \\
        0 & 0 & 0 & 0 & 0 & 0 & 0 & 0 & 0 & 0 & 0 & 0 & \lambda_{H} & 0 & 0 \\
        0 & 0 & 0 & 0 & 0 & 0 & 0 & 0 & 0 & 0 & 0 & 0 & 0 & \lambda_{HS} & 0 \\
        0 & 0 & 0 & 0 & 0 & 0 & 0 & 0 & 0 & 0 & 0 & 0 & 0 & 0 & \lambda_{H}
    \end{pmatrix}$ }
\end{equation}

All these above steps are encapsulated in the following function, which returns all the distinct eigenvalues:
\begin{lstlisting}[language=Mathematica]
Scan[Print, 
 EigenvaluesScatteringMatrix[
   VSinglet, assumption]["AllDistinctEigenvalues"]]
\end{lstlisting}
\begin{align}
        & \frac{1}{2}\left(3 \lambda_H + 6 \lambda_S - \sqrt{(3 \lambda_H - 6 \lambda_S)^2+4 \lambda_{HS}^2 }\right), \\
    & \frac{1}{2}\left(3 \lambda_H + 6 \lambda_S + \sqrt{(3 \lambda_H - 6 \lambda_S)^2+4 \lambda_{HS}^2 }\right),\\
    &\lambda_{HS},\\
    &\lambda_H.
\end{align}

\subsection{Example 2: Two-Higgs-doublet model}

Let us now work with an extensively studied model, the two-Higgs-doublet model (2HDM).

\begin{align}
V_4=&~\frac{\lambda_1}{2} \left(H_1^\dagger H_1\right)^2+\frac{\lambda_2}{2} \left(H_2^\dagger H_2\right)^2+\lambda_3 \left(H_1^\dagger H_1\right) \left(H_2^\dagger H_2\right)\nonumber\\
&+\lambda_4 \left(H_1^\dagger H_2\right) \left(H_2^\dagger H_1\right)+\bigg\{\frac{1}{2} \lambda_5 \left(H_1^\dagger H_2\right)^2\nonumber\\
&+\left[\lambda_6 \left(H_1^\dagger H_1 \right)  + \lambda_7 \left(H_2^\dagger H_2\right)\right] \left(H_1^\dagger H_2\right) +\text{h.c}\bigg\},\nonumber\\
    &\text{with}\quad H_{1,2}=\frac{1}{\sqrt{2}}\begin{pmatrix}
    \omega_1^{1,2}+i~\omega_2^{1,2} \\
    h_{1,2}+i~z_{1,2}\end{pmatrix}
\end{align}

The quartic part of the tree-level potential can be expressed in terms of real fields, while indicating which quartic couplings are real, with the following lines of code:
\begin{lstlisting}[language=Mathematica]
V4= \[Lambda]1/2 (H1\[Conjugate].H1)^2+\[Lambda]2/2 (H2\[Conjugate].H2)^2+\[Lambda]3(H1\[Conjugate].H1)(H2\[Conjugate].H2)+\[Lambda]4  (H1\[Conjugate].H2)(H2\[Conjugate].H1)+1/2 (\[Lambda]5 (H1\[Conjugate].H2)^2+ \[Lambda]5\[Conjugate] (H2\[Conjugate].H1)^2)+(\[Lambda]6 (H1\[Conjugate].H1) (H1\[Conjugate].H2)+\[Lambda]6\[Conjugate](H1\[Conjugate].H1) (H2\[Conjugate].H1))+(\[Lambda]7 (H2\[Conjugate].H2) (H1\[Conjugate].H2)+\[Lambda]7\[Conjugate] (H2\[Conjugate].H2) (H2\[Conjugate].H1));
\end{lstlisting}

\begin{lstlisting}[language=Mathematica]
V2HDM=ComplexExpand[V4/.{H1->{(\[Omega]11+I \[Omega]12)/Sqrt[2],(h1+I z1)/Sqrt[2]},H2-> {(\[Omega]21+I \[Omega]22)/Sqrt[2],(h2+I z2)/Sqrt[2]}},{\[Lambda]5,\[Lambda]6,\[Lambda]7}];
assumption = {{\[Lambda]1,\[Lambda]2,\[Lambda]3,\[Lambda]4}\[Element]Reals};\end{lstlisting}

\subsubsection{$\lambda_6 = \lambda_7 = 0$}

In the $\mathbb{Z}_2$-symmetric 2HDM, we can easily obtain the eigenvalues analytically:
\begin{lstlisting}[language=Mathematica]
Scan[Print, EigenvaluesScatteringMatrix[V2HDM /. {\[Lambda]6 -> 0, \[Lambda]7 -> 0}, assumption]["AllDistinctEigenvalues"]]
\end{lstlisting}

\begin{align}
    &\frac{1}{2}\left(3(\lambda_1+\lambda_2)\pm\sqrt{9(\lambda_1-\lambda_2)^2+4(2\lambda_3+\lambda_4)^2}\right),\\
    &\frac{1}{2}\left(\lambda_1+\lambda_2\pm\sqrt{(\lambda_1-\lambda_2)^2+4\lambda_4^2}\right),\\
    &\frac{1}{2}\left(\lambda_1+\lambda_2\pm\sqrt{(\lambda_1-\lambda_2)^2+4\vert\lambda_5\vert^2}\right),\\
    &\lambda_3+2\lambda_4\pm3\vert\lambda_5\vert,\\
    &\lambda_3\pm\vert\lambda_5\vert,\\
    &\lambda_3\pm\lambda_4.
\end{align}

\subsubsection{$\lambda_6 \neq 0, \lambda_7 = 0$ and $\lambda_6 \neq 0, \lambda_7 \neq 0$}

When the $\lambda_6$ term is considered, it is no longer possible to obtain the eigenvalues analytically, because block-diagonalised scattering matrix is made of two large blocks, $B_{20}$ and $B_{16}$.

\begin{lstlisting}[language=Mathematica]
sm = ScatteringMatrix[V2HDM/.\[Lambda7]->0, assumption]["ScatteringMatrix"];
smBlock = MakeBlockDiagonal[sm];
Length /@ smBlock["Components"]
\end{lstlisting}
\begin{equation}
    \{20,~16\}
\end{equation}

 The structure of that matrix is the following:\newline
 
\begin{lstlisting}[language=Mathematica]
ArrayPlot@SparseArray[smBlock["BlockMatrix"]]
\end{lstlisting}
\begin{figure}[!h]
\centering
\includegraphics[scale=0.5855,trim = 0cm 0cm 0cm 0cm, clip]{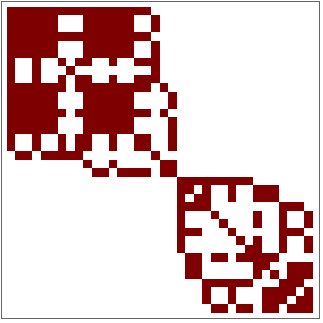}
\end{figure}

Therefore in that case, eigenvalues should be computed numerically.

When both $\lambda_6$ and $\lambda_7$ terms are considered, we again obtain the blocks $B_{20}$ and $B_{16}$ but they are now less sparse. Obviously in that case, it is even more difficult to obtain the eigenvalues analytically and they should thus be computed numerically.

\bibliographystyle{bibi}
\bibliography{biblio.bib}

\end{document}